\documentclass[aps,prl,twocolumn,superscriptaddress]{revtex4-1} 
\usepackage{graphicx}
\usepackage{xcolor}
\usepackage{subfigure}
\usepackage{mathtools}
\usepackage{physics,amsmath}
\usepackage{epsfig}
\usepackage{chemfig}
\usepackage[colorlinks= true,urlcolor=blue,linkcolor= blue,citecolor=blue,bookmarks=false,pdfstartview=]{hyperref}
\usepackage{siunitx}
\usepackage{bm} 
\usepackage{xr}
\usepackage{xcolor}

\begin{document}

\title{Direct evidence from high-field magnetotransport for a dramatic change of quasiparticle character in van der Waals ferromagnet Fe$_{3-x}$GeTe$_2$}

\author{S. Vaidya}
 \email{s.vaidya@warwick.ac.uk}
\affiliation{Department of Physics, University of Warwick, Gibbet Hill Road, Coventry, CV4 7AL, UK}\author{M. J. Coak}
\affiliation{Department of Physics, University of Warwick, Gibbet Hill Road, Coventry, CV4 7AL, UK}
\affiliation{School of Physics \& Astronomy, University of Birmingham, Edgbaston, Birmingham, B15 2TT, UK}
\author{D. A. Mayoh}
\affiliation{Department of Physics, University of Warwick, Gibbet Hill Road, Coventry, CV4 7AL, UK}
\author{M. R. Lees}
\affiliation{Department of Physics, University of Warwick, Gibbet Hill Road, Coventry, CV4 7AL, UK}
\author{G. Balakrishnan}
\affiliation{Department of Physics, University of Warwick, Gibbet Hill Road, Coventry, CV4 7AL, UK}
\author{J. Singleton}
\affiliation{National High Magnetic Field Laboratory (NHMFL), Los Alamos National Laboratory, Los Alamos, NM, USA}
\author{P. A. Goddard}
 \email{p.goddard@warwick.ac.uk}
\affiliation{Department of Physics, University of Warwick, Gibbet Hill Road, Coventry, CV4 7AL, UK}

\begin{abstract}

Magnetometry and magnetoresistance (MR) data taken on the van der Waals ferromagnet Fe$_{3-x}$GeTe$_2$ (FGT) reveal three distinct contributions to the MR: a linear negative component, a contribution from closed Fermi-surface orbits, and a $H^2$ enhancement linked to a non-coplanar spin arrangement. Contrary to earlier studies on FGT, by accounting for the field dependence of the anomalous Hall effect, we find that the ordinary Hall coefficient decreases markedly below 80\,K, indicating a significant change in character of the electrons and holes on the Fermi surface at this temperature. The resulting altered ground state eventually causes the Hall coefficient to reverse sign at 35\,K. Our Hall data support the proposal that Kondo-lattice behavior develops in this $d$-electron material below 80 K. Additional evidence comes from the negative linear component of the MR, which arises from electron-magnon scattering with an atypical temperature dependence attributable to the onset of Kondo screening.  

\end{abstract}
\maketitle

Magnetic van der Waals (vdW) materials are emerging as an exciting testing ground for exploring fundamental theories of magnetism in the extreme two-dimensional (2D) limit~\cite{Ajayan_2016_review, burch_2018_review, khan2020, Wang2022magnetic_genome}, while the ease with which they can be exfoliated and reassembled into varied heterostructures makes them promising ingredients in devices that harness the interplay between magnetic and electronic degrees of freedom~\cite{Liang_2019_review, Sierra_2021_review}. Among these, the Fe-based compounds, such as Fe$_{3-x}$GeTe$_2$ (FGT) and Fe$_3$GaTe$_2$, have received significant attention due to their high Curie temperatures ($T_{\text{C}}$) and metallic behavior seldom seen in ferromagentic (FM) vdW materials. FGT in particular, with maximum reported $T_{\text{C}}$ of $230$\,K~\cite{Deiseroth2006}, has been the subject of numerous studies exhibiting a wide range of phenomena, including topological nodal lines~\cite{kim2018}, a topological Hall effect~\cite{Wang2017, Chowdhury2021}, competing AFM fluctuations~\cite{Yi2016, bai2022, Bao2022} and skyrmions~\cite{Li2022}. 

The uniqueness of FGT in this regard has led to the development of multiple heterostructures that demonstrate potential for low-power spintronic technologies~\cite{albarakati2019, Wu2020, Wu2022}. Nevertheless, consensus is yet to be reached on the fundamental magnetic and electronic processes that underpin these devices. The itinerant versus localised nature of the magnetism is of particular interest, and some recent angle-resolved photoemission spectroscopy (ARPES) and neutron-scattering studies suggest that FGT displays a coexistence of the two, with resulting Kondo-lattice behavior developing at low temperatures~\cite{Zhang2018a, Bao2022}. However, whilst Kondo-lattice behavior is widely accepted in $f$-electron systems, its presence in $3d$ compounds is more controversial~\cite{liu2020_CaCu3Ru4O12,Yang2022_Kondo_rev}. While some degree of electronic correlation is known to exist in FGT (estimated quasiparticle effective masses range between 1.6 and 15.9 times the bare electron mass~\cite{Zhu2016, Sun2023, Zhang2018a, Trainer2022}), strong evidence for Kondo screening in the transport properties is so far elusive. 

 \begin{figure}[t]
     \includegraphics[width= \linewidth]{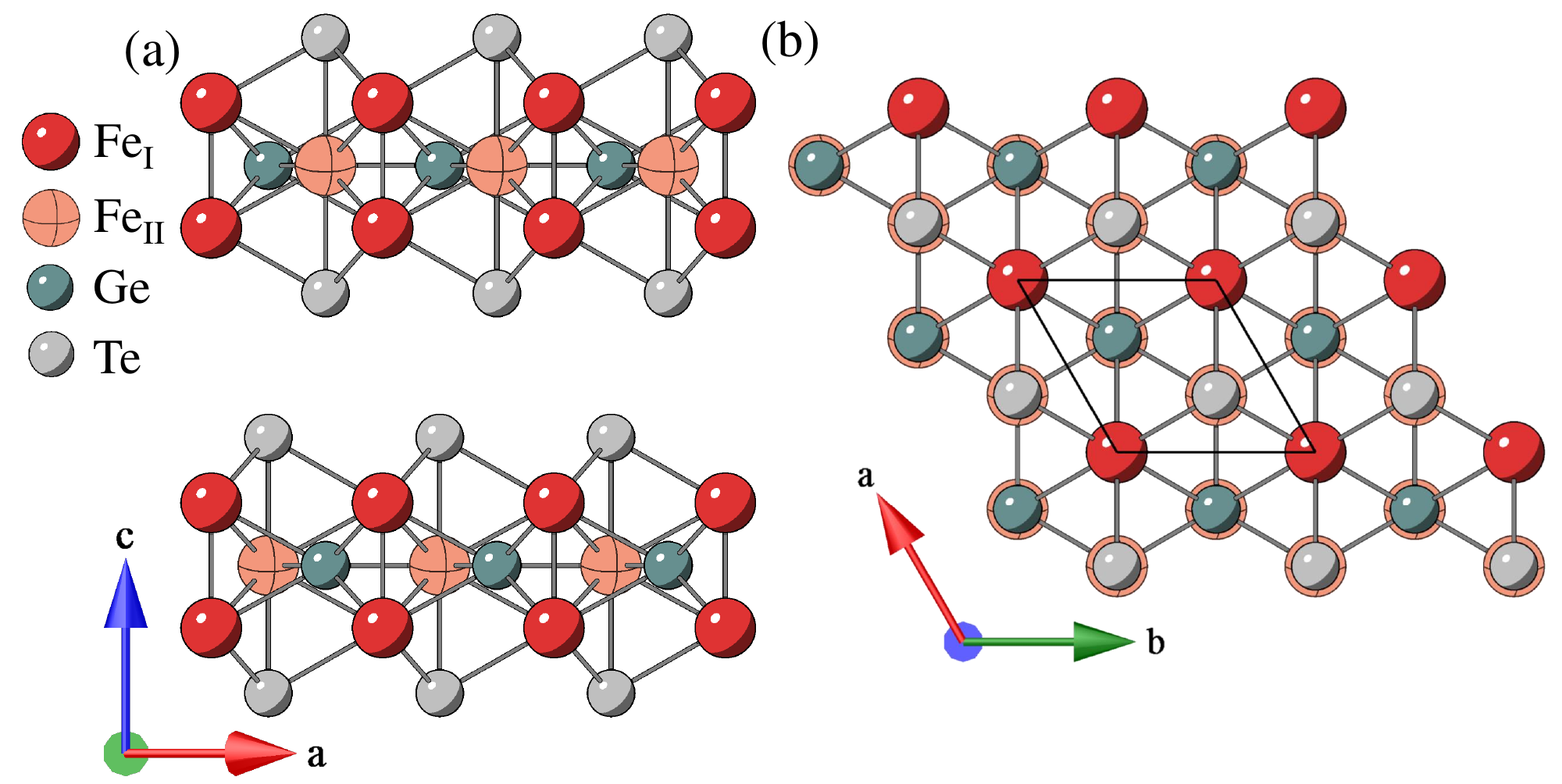}
     \caption{Crystal structure (from data in Ref.~\cite{Villars2023:CIF}) of Fe$_3$GeTe$_2$ showing the Fe$_{\rm I}$ and Fe$_{\rm II}$ sites. View along (a) the $b$-axis within the vdW-bonded layers, and (b) the $c$-axis, perpendicular to the layers.}
     \label{fig: Struct}
 \end{figure}

\begin{figure}[t]
    \includegraphics[width= \linewidth]{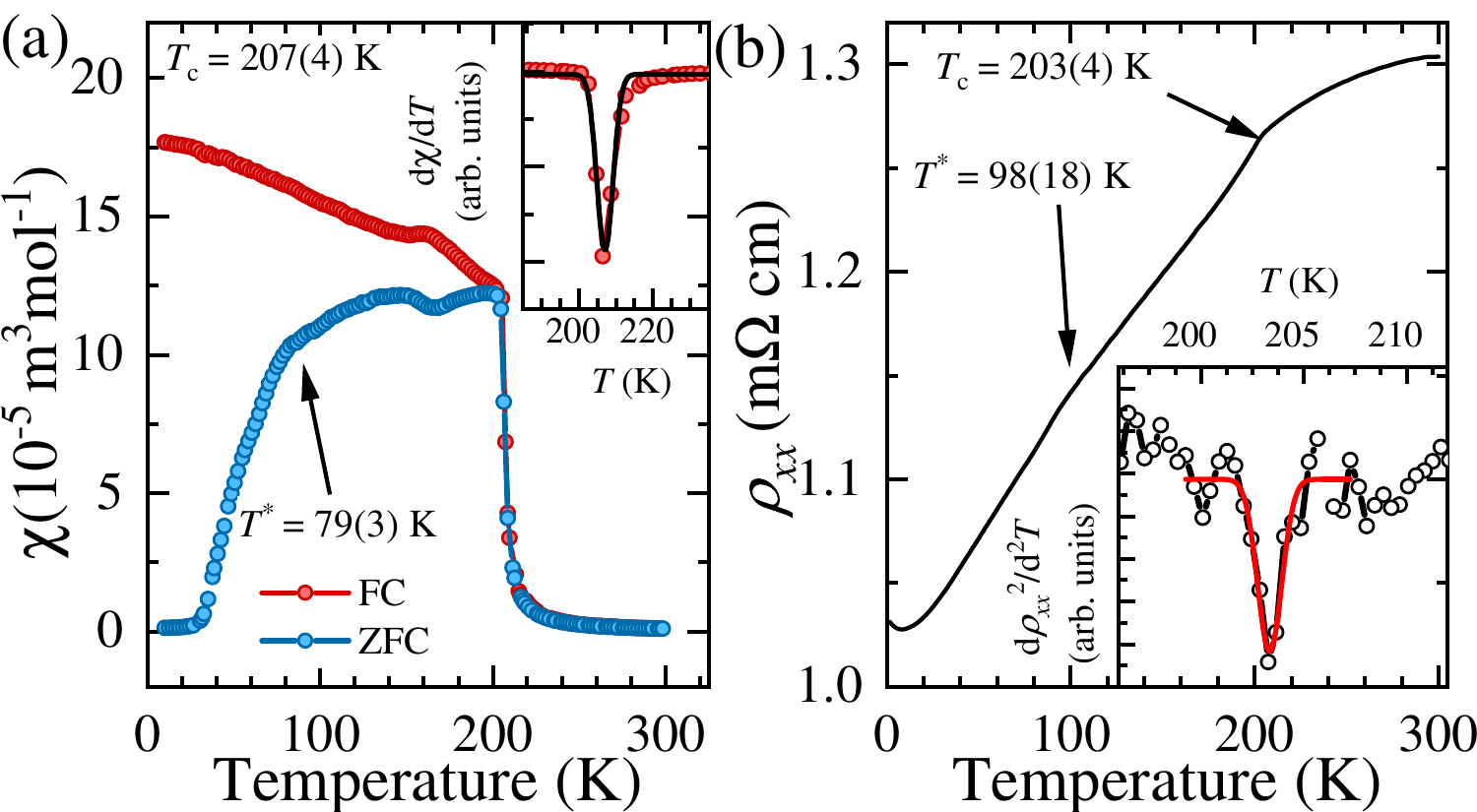}
    \caption{(a) Zero-field-cooled (blue) and field-cooled (red)  temperature-dependent magnetic susceptibility $\chi(T)$ for a single crystal of Fe$_{3-x}$GeTe$_2$, with a field of $0.01$\,T applied parallel to the $c$-axis. Inset: A dip in $\text{d}\chi\slash\text{d}T$ indicates the ferromagnetic transition at $T_{\rm C}$. (b) Temperature dependence of the in-plane resistivity $\rho_{xx}$ in zero applied field. Inset: A dip in $\text{d}^{2}\rho_{xx}\slash\text{d}T^{2}$ occurs at $T_{\rm C}$.}
    \label{fig: Characterisation}
\end{figure}

One reason that these issues remain unresolved is the paucity of high-quality transport data on samples close to the desired stoichiometry of Fe$_3$GeTe$_2$. The synthesis of this material is known to be challenging due to a propensity for vacancies on the Fe$_{\rm I}$ site (see Fig.~\ref{fig: Struct}) that suppress the FM ordering from the ideal value of $T_{\text{C}} = 230$\,K~\cite{May2016, Mayoh2021}. Several transport studies have been performed on samples exhibiting $T_{\text{C}}$ less than 160\,K~\cite{Liu_2018, You2019, Ke2020}, indicating that more than one in eight Fe ions are missing from the structure~\cite{Mayoh2021}, with detrimental repercussions for the transport properties. Transport measurements on higher quality samples do exist~\cite{Wang2017, Saha2023}, however, these papers overlook the significant field dependence of the factors contributing to the anomalous Hall effect (described below) when determining the behavior of the ordinary Hall coefficient. By performing magnetometry and magnetotransport measurements in high magnetic fields, we are able to take into account the presence of a non-saturating magnetization, as well as the marked field dependence of the longitudinal magnetoresistance. In this way, we uncover evidence of multiple charge carrier types and a temperature-dependent evolution of the Fermi surface (FS) properties that supports the Kondo-lattice scenario proposed~\cite{Zhang2018a, Bao2022} in FGT.

Large single crystals of FGT are grown using the chemical-vapour-transport method described in Ref.~\cite{Mayoh2021}. Bulk magnetometry measurements up to 9\,T, are carried out using a vibrating sample magnetometer and measurements up to 7\,T using an Magnetic Property Measurement System-XL SQUID magnetometer. Pulsed-field magnetization measurements are performed at the National High Magnetic Field Laboratory, Los Alamos, using several coaligned single crystals. Longitudinal transport and Hall-effect measurements, up to 9\,T, are carried out in a Quantum Design Physical Property Measurement System and high-field transport measurements are carried out at Los Alamos on flat-plate samples. Further experimental details including a discussion of demagnetization corrections are given in the Supplemental Material~\cite{supplementary}.

\begin{figure}[t]
    \hspace{-1em}%
    \begin{subfigure}
        \centering
        \includegraphics[width= 0.5\linewidth]{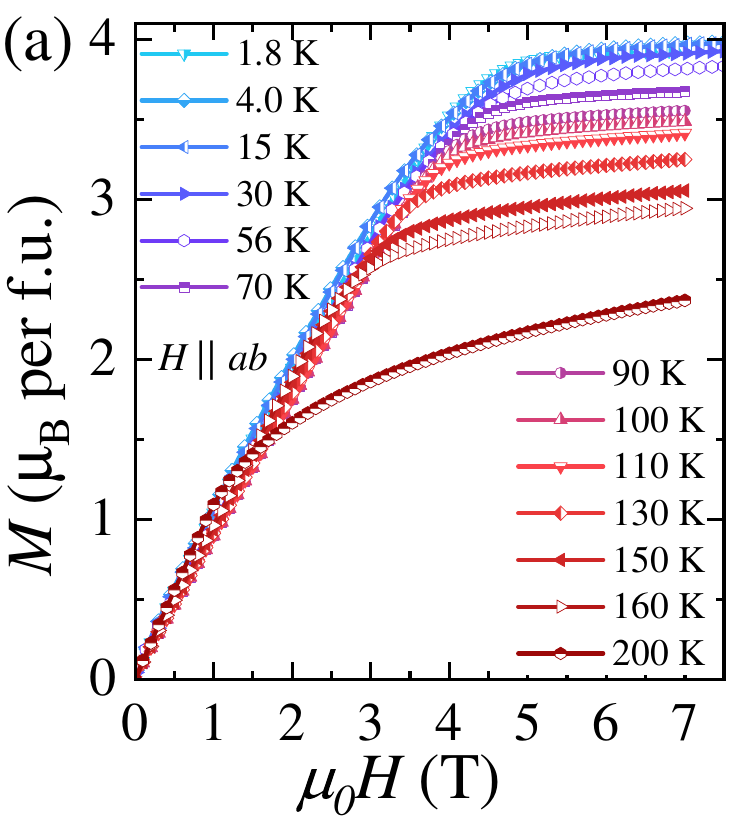}
    \end{subfigure}
    \hspace{-1em}%
    \begin{subfigure}
        \centering
        \includegraphics[width= 0.5\linewidth]{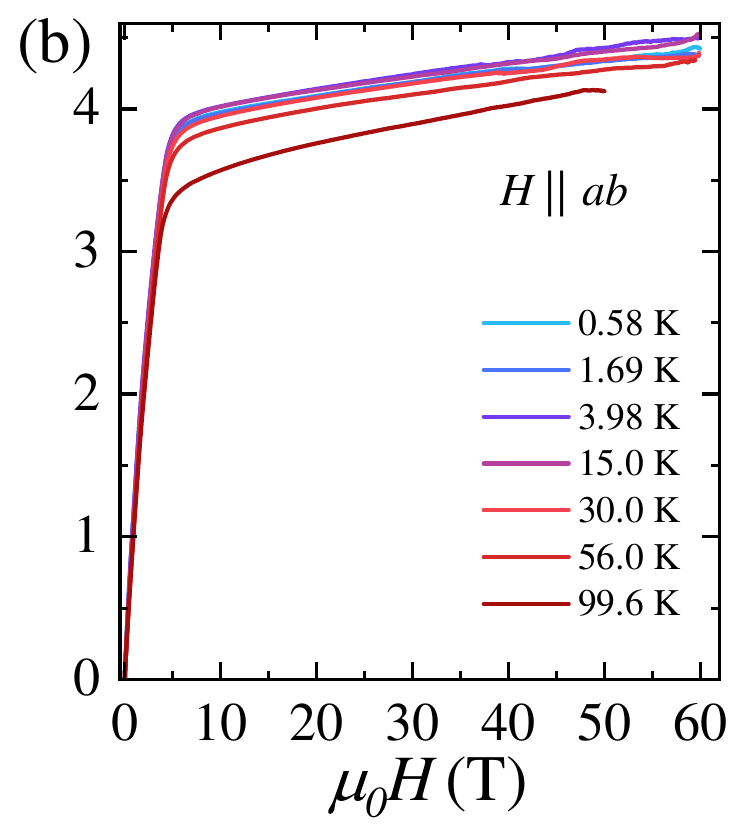}
    \end{subfigure}
    \begin{subfigure}
        \centering
        \includegraphics[width= 0.5\linewidth]{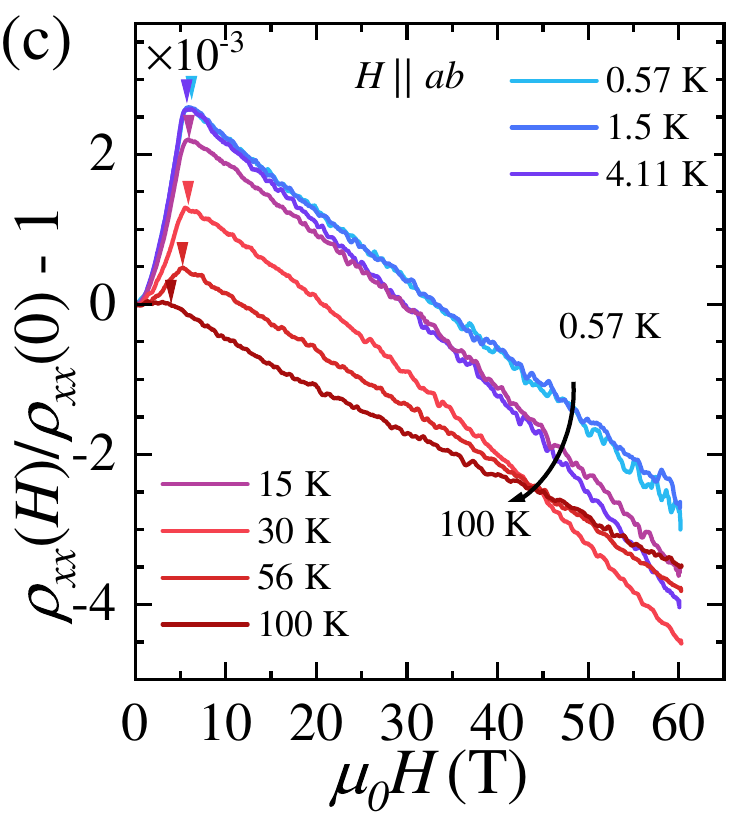}
    \end{subfigure}
    \hspace{-1em}%
    \begin{subfigure}
        \centering
        \includegraphics[width= 0.5\linewidth]{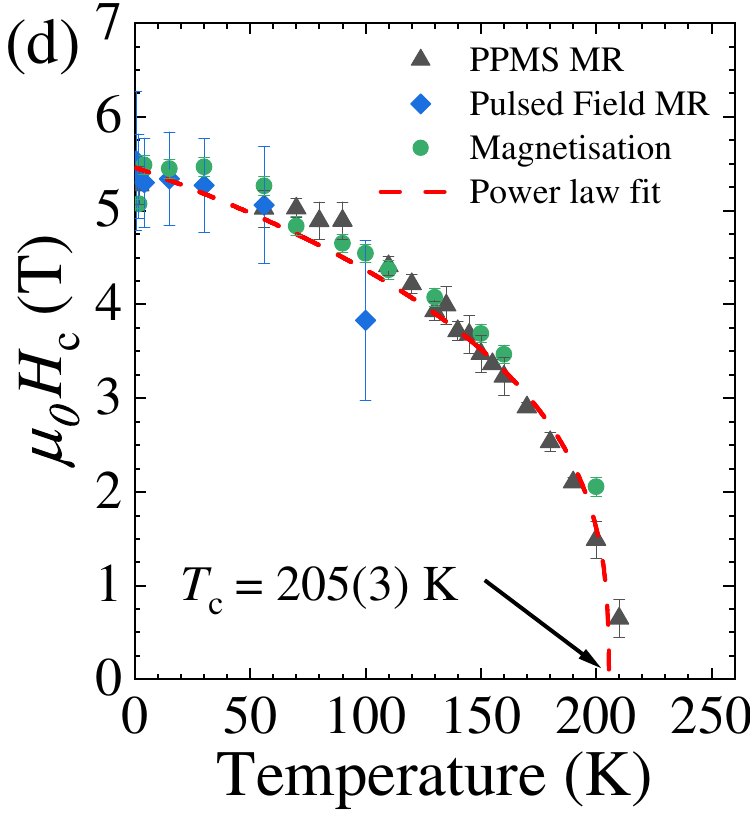}
    \end{subfigure}
    \caption{Measurements with the applied field $H\parallel ab$. (a) Magnetization up to $7$\,T, (b) pulsed-field magnetization and (c) longitudinal in-plane magnetoresistance. (d) $H_{ab}$---$T$ phase diagram. Black triangles and blue diamonds denote the $H_{\rm c}$ feature discussed in the text and marked by arrows in (c). Green circles mark the apparent saturation seen in the magnetization. Red dashed line is a fit to a power law described in the text.
    \label{fig: H_parra_ab}
    }
\end{figure} 
\begin{figure*}[ht]
\includegraphics[width= \linewidth]{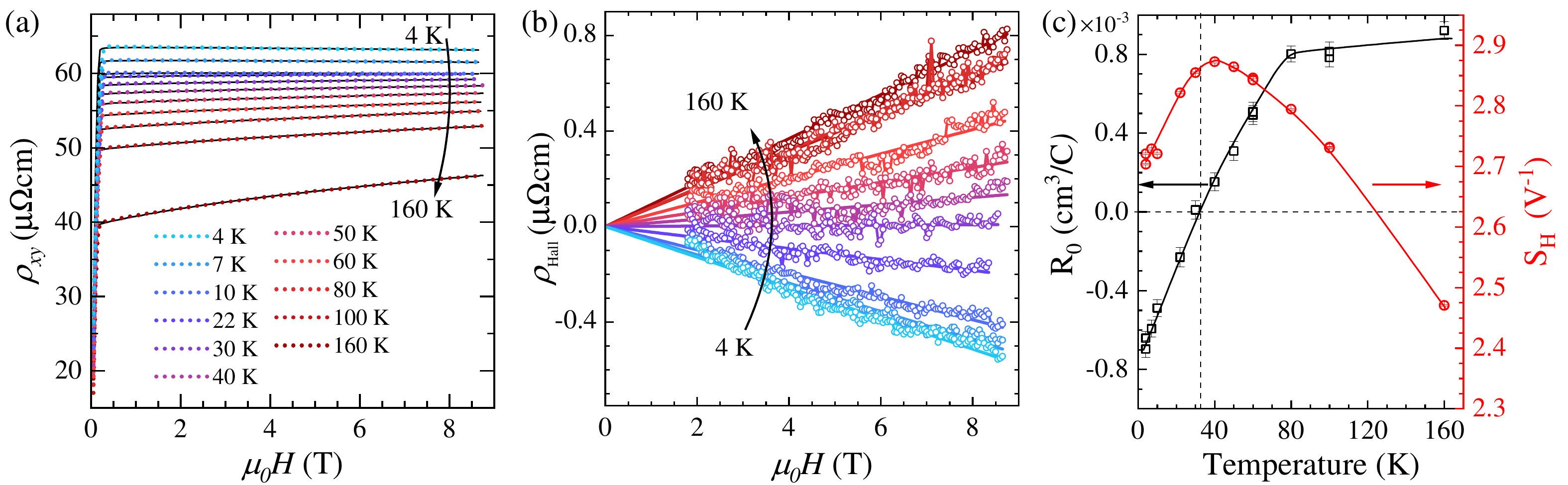}
\caption{(a) Measured in-plane Hall resistivity $\rho_{xy}$ (circles) as a function of $H\parallel c$. The solid lines are the result of fits described in the text. For clarity, the curves have been offset by $2$\,$\mu\Omega\text{cm}$. (b) The ordinary Hall effect component $\rho_{\text{Hall}}$ extracted from the fitting process. (c) Temperature dependence of the ordinary Hall coefficient $R_{0}$ (black squares) and the AHE scale factor $S_{H}$ (red circles).Solid lines are a guide to the eye.}
\label{fig: Hall}
\end{figure*}

Field-cooled (FC) and zero-field-cooled (ZFC) temperature-dependent magnetic susceptibilities of FGT were measured with an applied field of $\mu_{0}H = 0.01$\,T and shown in Fig.~\ref{fig: Characterisation}(a). A sharp deviation from Curie-Weiss behavior and bifurcation of the two curves indicate FM ordering at $T_{\text{C}} = 207(4)$\,K. The suppression of $T_{\text{C}}$ from $230$\,K suggests a Fe deficiency of $x \approx 0.15$ in the sample~\cite{Mayoh2021}. Energy-dispersive X-ray analysis confirms the sample composition to be Fe$_{2.90}$GeTe$_{2.12}$ ($x = 0.10(1)$), which is similar to the best samples for which transport and magnetometry data exist~\cite{Wang2017,Chowdhury2021,Saha2023}. A kink is seen in both the FC and ZFC curves at $T = 161(1)$\,K, consistent with previous reports of possible AFM fluctuations~\cite{Yi2016}. Furthermore, there is a sudden drop in the ZFC on cooling data at $T^{*} = 79(3)$\,K. 

Figure~\ref{fig: Characterisation}(b) displays the temperature dependence of the in-plane longitudinal resistivity, $\rho_{xx}$. The curves exhibit metallic behavior with a sharp change in gradient, observed as a minimum in $d\rho^{2}_{xx}/d^{2}T$, associated with the FM ordering at $T_{\text{C}} = 203(4)$\,K. This is in agreement with the $\chi(T)$ data. An additional change in the gradient is observed at $T^{*} = 98(18)$\,K, prior to an upturn below $T = 8.0(3)$\,K. The characteristic temperature $T^{*}$ has previously been suggested to signify the onset of Kondo-lattice behavior in FGT~\cite{Zhang2018a, Bao2022}.

Figures~\ref{fig: H_parra_ab}(a) and 3(b) present the isothermal magnetization $M$ at various temperatures as a function of $H\parallel ab$. $M(H)$ rises quickly at low fields to an apparent saturation, which is followed by a much slower increase up to 60\,T, likely due to the movement of domain walls stuck on strong pinning sites.

The longitudinal in-plane magnetoresistance (MR) with $H\parallel ab$ (with an in-plane current $I\perp H$) is displayed in Fig.~\ref{fig: H_parra_ab}(c). At low fields and low temperatures, the MR increases $\propto H^{2}$, before reversing gradient abruptly at $H_{\text{c}}$. By $80$\,K, the low-field $H^2$ dependence of the MR is no longer present; however, a shoulder feature at $H_{\text{c}}$ persists up to $T_{\text{c}}$ as shown in the Supplemental Material~\cite{supplementary}. Figure~\ref{fig: H_parra_ab}(d) shows the result of plotting $H_{\rm c}$ as a function of temperature alongside the apparent saturation field seen in the $M(H\parallel ab)$ data. It can be seen that the feature found in the transport measurement maps out the same phase boundary seen in the in-plane magnetometry. A power-law fit of the form $H_{\text{c}} = H_{\text{0}}(1 - T/T_{\text{c}})^{\delta}$ yields $\mu_{0}H_{\text{0}} = 5.46(7)$\,T, $\delta = 0.33(2)$ and $T_{\text{C}} = 205(3)$\,K, in excellent agreement with the FM ordering temperature determined earlier. A similar phase boundary has previously been mapped through observations of the topological Hall effect, which arises due to the non-coplanar spin configuration that exists for $H\parallel ab$ below the saturation field~\cite{Wang2017, Chowdhury2021, You2019}. However, an enhancement of the longitudinal MR associated with the canted phase was not seen in those studies (likely for the reasons discussed earlier). We note that the MR behavior is symmetric in field and so does not arise from a poor subtraction of the Hall component of resistivity~\cite{supplementary}. Nevertheless, quite why the MR would show a $H^2$ increase within the non-coplanar magnetic phase remains to be explained.

\begin{figure}[t]
    \includegraphics[width= 0.9\linewidth]{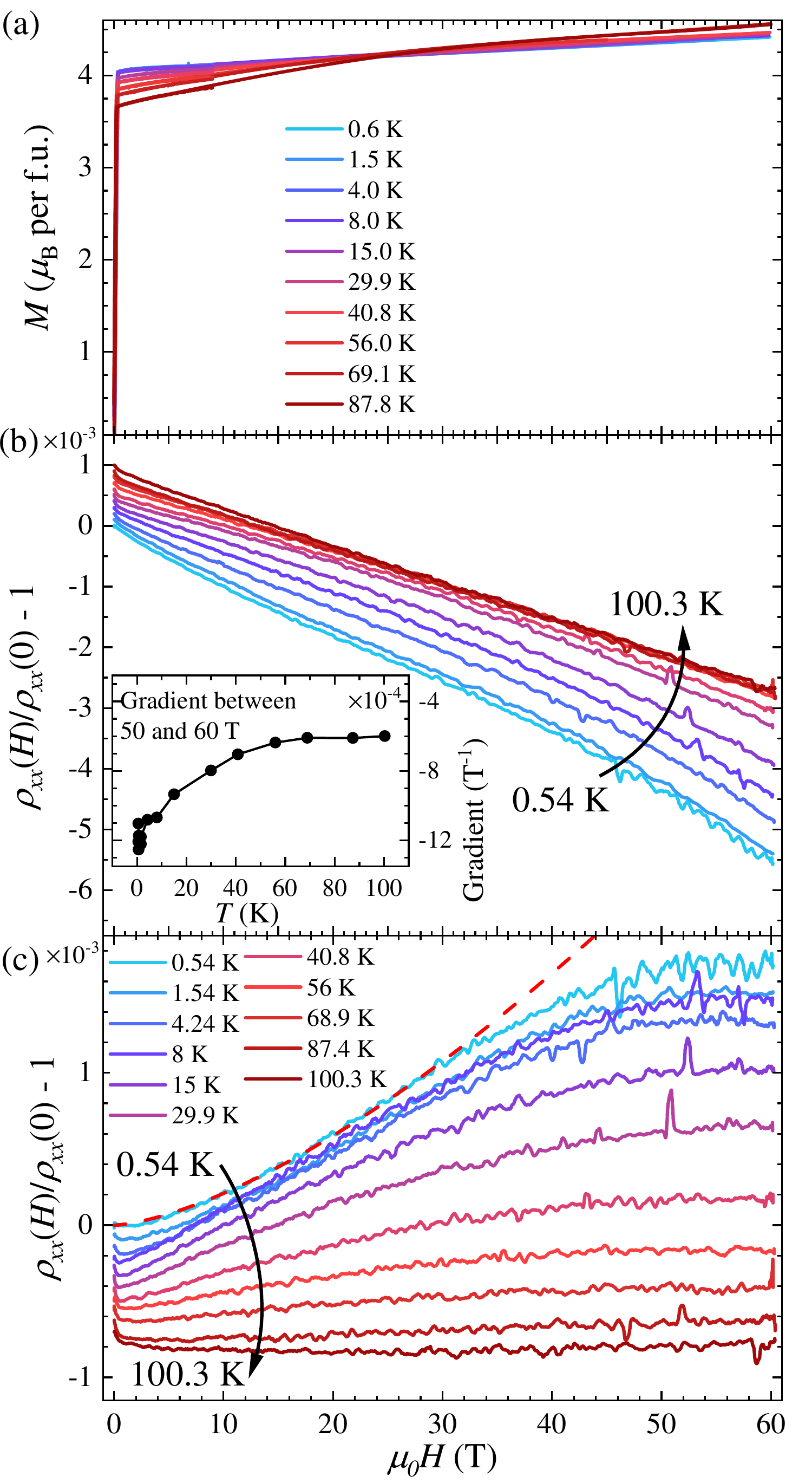}
    \caption{Measurements with $H\parallel c$. (a) Magnetization up to $60$\,T. (b) In-plane magnetoresistance (MR). An offset of $0.1\%$ has been applied for clarity. Inset: Gradient of the negative linear component. (c) MR after the linear component has been removed (curves offset for clarity). The red dashed line shows an $H^{1.5}$ fit to the $0.54$\,K data.}
    \label{fig: Parra_c}
\end{figure}

Figure~\ref{fig: Hall}(a) shows the Hall resistivity $\rho_{xy}$ (circles) measured with $I\parallel ab$ and $H\parallel c$. Typically in FM conductors, the Hall effect is composed of two components
\begin{equation}
    \rho_{xy} = \rho_{\text{Hall}} + \rho_{\text{AH}} =  R_{0}\mu_{0}H + \mu_{0}R_{\text{S}}M,
\end{equation}
where the $\rho_{\text{Hall}}$ is the ordinary Hall effect with coefficient $R_{0}$ and $\rho_{\text{AH}}$ is the anomalous Hall effect (AHE) with coefficient $R_{\text{S}}$. Previous studies have shown that the Karplus-Luttinger mechanism, with $R_{\text{S}} \propto \rho_{xx}^2$, is dominant in FGT~\cite{Wang2017, kim2018, Saha2023}. Consequently, the Hall resistivity can be written as
\begin{equation}
    \label{eq: hall_KL}
    \rho_{xy} = R_{0}\mu_{0}H + S_{\text{H}}\rho^{2}_{xx}M,
\end{equation}
where $S_{\text{H}}$ is the AHE scaling factor.  In this equation, the field-dependence of both $M$ and $\rho_{xx}$ is neglected in earlier reports on FGT. Figure~\ref{fig: Parra_c}(a) and 5(b), respectively, show the results of our measurements of these quantities for $H\parallel c$. Similar to the case for an in-plane field, the $M(H)$ data with $H\parallel c$ display a sharp rise to an apparent saturation, but then continue to increase more gradually up to $60$\,T. 

The MR data in Fig.~\ref{fig: Parra_c}(b) exhibit a strong negative trend up to the highest fields. Our data show that these contributions to the AHE are not insignificant and must be taken into account to achieve a reliable analysis of the Hall data.  If this is done, then the parameters $S_{\text{H}}$ and $R_{0}$ in Eq.~\ref{eq: hall_KL} can be extracted from linear fits to $\rho_{xy}/\mu_{0}H$ vs $\rho^{2}_{xx}M/\mu_{0}H$, at $\mu_{0}H \geq 2$\,T. The parameters extracted in this way can be inserted back into Eq.~\ref{eq: hall_KL} as a check, and the  calculated $\rho_{xy}$ is seen to be in excellent agreement with the raw data (see solid lines in Fig.~\ref{fig: Hall}(a)). Moreover, the isolated $\rho_{\text{Hall}}$ component is shown in Fig.~\ref{fig: Hall}(b) and is found to be linear in $H$ at all temperatures, as expected.

The temperature dependence of the extracted $R_{0}$ and $S_{\text{H}}$ parameters are shown in Fig.~\ref{fig: Hall}(c). Between $160$ and $80$\,K, $R_{0}$ is roughly constant and positive, indicating the presence of dominant hole carriers. On cooling further, the Hall coefficient drops steadily, becoming negative for $T\lesssim 35$\,K. $S_{\text{H}}$ exhibits a maximum at around the same temperature. The behavior of $R_0(T)$ directly confirms the presence of both electron and hole pockets in the FS of FGT, as suggested by previous ARPES measurements~\cite{Zhang2018a, Xu2019}. 

The sudden drop in $R_0(T)$ at around 80\,K implies that a significant change in the FS takes place at this temperature. A previous report comparing neutron scattering and dynamical mean-field theory suggested that an orbitally selective Mott transition occurs at $T \approx 100$\,K~\cite{bai2022}, which would indeed result in a marked change in $R_0$. However, the transition predicted in Ref.~\cite{bai2022} would lead to a loss of an electron pocket at the $K$-point. By contrast, our results suggest that a significant increase in electron density and/or mobility (relative to the holes) sets in at 80\,K \footnote{To put it another way, a significant decrease in the density and/or mobility of the holes, relative to those of the electrons sets in at $80$\,K}. On the other hand, the abrupt drop in $R_{0}$ occurs very close to the features we observe in $\chi(T)$ and $\rho_{xx}$ at $T^{*}=80$\,K, and which have been seen by other authors. Using evidence from ARPES, STM and neutron spectroscopy, Refs.~\cite{Zhang2018a} and~\cite{Bao2022} suggest that $T^{*}$ marks the onset temperature of Kondo-lattice behavior, below which coherent heavy quasiparticles emerge and act to screen the local moments, as is seen in $f$-electron heavy-fermion materials. The resulting alteration in the FS properties would readily explain the drop we see in the Hall coefficient below $T^*$. 

Further support for the Kondo-lattice picture comes from the $H\parallel c$ MR data shown in Fig.~\ref{fig: Parra_c}(b), which exhibit a strong negative trend for all temperatures up to 60\,T [similar to the in-plane field data above $H_{\rm c}$, see Fig.~\ref{fig: H_parra_ab}(c)]. The absence of the low-field $H^2$ dependence in the $H\parallel c$ MR data confirms that that behavior is indeed connected to the canted magnetic phase which only exists for in-plane fields. For the $H\parallel c$ data, the MR is dominated by a negative linear component, with a smaller additional dependence apparent at low temperatures and fields. Fitting the MR between 50 and 60\,T to obtain the negative linear component and subtracting it from the data yields a positive, saturating component that diminishes as temperature is raised and disappears completely close to 100\,K, as shown in Fig.~\ref{fig: Parra_c}(c). 

Positive MR is typically caused by the orbital motion of charge carriers on a non-spherical FS. A simple closed FS pocket is expected to produce a saturating MR with a $H^2$ dependence at low-fields, while a non-saturating MR arises from either open FS orbits or perfect charge compensation in a simple two-band model (see e.g. Ref.~\cite{Zhang2019_MR_FS}). In our data at $0.54$\,K, the MR below $30$\,T can be described by $\text{MR} \propto H^{\delta}$, with $\delta = 1.5(1)$, before saturating at ${H}_{\text{app}} \approx 50$ T. Such behavior is consistent with the presence of different carrier types on multiple closed FS pockets and a marked deviation from charge compensation, a situation that is in agreement with the $R_0$ data presented earlier. It is also in keeping with prior theoretical and experimental work on FGT, which consistently reveals multiple closed orbits within the $k_{x} - k_{y}$ plane~\cite{Zhang2018a, Trainer2022, bai2022}. 

As mentioned, at all temperatures measured, the MR is dominated by a negative linear component. Such a response is well established in FMs 
as the result of a reduction in electron-magnon scattering caused by the damping of spin waves with applied field~\cite{Raquet2002}, a situation that has been suggested for FGT previously~\cite{Ke2020, Saha2023}. Typically, because the magnon population diminishes at low temperatures (where it is harder to flip a spin), the strength of the negative MR is found to decrease on cooling~\cite{Raquet2002}. However, the situation in FGT might be expected to be different. The inelastic neutron scattering data presented in Ref.~\cite{Bao2022} show that, in a departure from the typical behavior of spin waves, the low-energy magnons in FGT are more coherent at elevated temperatures, a result that is attributed to enhanced Kondo screening of local moments as the temperature is reduced below $T^*$. As a result, one might expect that the additional damping caused by an applied magnetic field would have a larger effect on electron-magnon scattering as the temperature is reduced, than at high temperatures where the lifetime and coherence of the spin waves are enhanced. Accordingly, we find that the negative MR gradient in FGT begins to increase as the temperature is reduced below 70\,K (see Fig.~\ref{fig: H_parra_ab}(b)), which is close to the $T^*\approx 80$\,K value seen in our measurements of $\chi(T)$, $\rho_{xx}(T)$ and $R_0(T)$.   

Taken together, our data paint a consistent picture of a clear change in the character (i.e., effective mass, scattering rates and possibly relative numbers) of the electrons and/or holes at the FS, caused by the onset of Kondo-lattice physics. Our results establish a connection between the transport properties and the earlier results of ARPES and neutron spectroscopy highlighting the heavy-fermion-like properties of FGT. The interplay of localised and itinerant magnetism, common in $f$-electron materials, is still an unusual proposition in transition-metal systems and warrants further investigation using other FS-sensitive probes.

\begin{acknowledgments}
This project has received funding from the European Research Council (ERC) under the European Union’s Horizon 2020 research and innovation programme (grant agreement No. 681260) and the UK Engineering and Physical Sciences Research Council (EPSRC) (grant no. EP/T005963/1 and EP/N032128/1). A portion of this work was performed at the National High Magnetic Field Laboratory (NHMFL), which is supported by National Science Foundation Cooperative Agreement Nos. DMR-1644779 and DMR-2128556 and the Department of Energy (DOE). J.S. acknowledges support from the DOE BES program “Science at 100 T,” which permitted the design and construction of much of the specialized equipment used in the high-field studies. For the purpose of open access, the author has applied a Creative Commons Attribution (CC-BY) licence to any Author Accepted Manuscript version arising from this submission. Data presented in this paper will be made available at XXXXXXX.

\end{acknowledgments}

\bibliography{main.bib}

\end{document}


\title{Supplementary Material accompanying \\ Direct evidence for dramatic change in quasiparticle character in van der Waals ferromagnet Fe$_{3-x}$GeTe$_2$ from high-field magnetotransport measurements}

\author{S. Vaidya}
\affiliation{Department of Physics, University of Warwick, Gibbet Hill Road, Coventry, CV4 7AL, UK}\author{M. J. Coak}
\affiliation{Department of Physics, University of Warwick, Gibbet Hill Road, Coventry, CV4 7AL, UK}
\affiliation{School of Physics \& Astronomy, University of Birmingham, Edgbaston, Birmingham, B15 2TT, UK}
\author{D. A. Mayoh}
\affiliation{Department of Physics, University of Warwick, Gibbet Hill Road, Coventry, CV4 7AL, UK}
\author{M. R. Lees}
\affiliation{Department of Physics, University of Warwick, Gibbet Hill Road, Coventry, CV4 7AL, UK}
\author{G. Balakrishnan}
\affiliation{Department of Physics, University of Warwick, Gibbet Hill Road, Coventry, CV4 7AL, UK}
\author{J. Singleton}
\affiliation{National High Magnetic Field Laboratory (NHMFL), Los Alamos National Laboratory, Los Alamos, NM, USA}
\author{P. A. Goddard}
\affiliation{Department of Physics, University of Warwick, Gibbet Hill Road, Coventry, CV4 7AL, UK}

\maketitle
\tableofcontents

\section{Sample Synthesis}

Single crystals of Fe$_{3-x}$GeTe$_2$ (FGT) are grown using the chemical vapour transport method. Stoichiometric quantities of Fe(STREM Chemicals, Inc., 99.99$\%$), Ge (Acros Organics, 99.999$\%$), and Te (AlfaAesar, 99.99$\%$) powder are sealed in evacuated quartz ampoule, with $5$\,mg/cm$^{3}$ of iodine as transport agent. A two-zone furnace is used to hold the source and sink end of the tube at $750\,^{\circ}$C and $675\,^{\circ}$C respectively, for $2$ weeks, before cooling to room temperature. The growth process results in the formation of silvery metallic platelet-shaped single crystals with areas of $\sim2\times2$\,mm$^{2}$ as previously reported in Ref.~\cite{Mayoh2021}. Energy-Dispersive X-ray spectroscopy (EDX) is conducted to determine a sample composition of Fe$_{2.90}$GeTe$_{2.12}$ ($x = 0.10(1)$). All measurements are conducted on samples from the same crystal growth batch.

\section{SQUID and VSM Magnetometry}

Temperature-dependent magnetic susceptibility, $\chi(T)$ measurements are performed in a Quantum Design MPMS XL SQUID magnetometer. Field-dependent magnetization, $M$, up to $7$\,T are also performed in the MPMS XL, while measurements up to $9$\,T are conducted in an Oxford Instruments vibrating sample magnetometer (VSM). For measurements of $M$, a single crystal sample measuring $1.95 \times 1.45 \times 0.68$\,mm$^3$ was orientated and affixed to a low-background PEEK sample holder using cyanoacrylate glue. For measurement with $H\parallel c$, a demagnetization correction of the form,
\begin{equation}
    H = H_{\text{app}} - DM(H_{\text{app}}),
\end{equation}
was applied. Here, $H$ is the internal field in the sample, $H_{\text{app}}$ is the applied magnetic field and the demagnetizing factor $D = 0.543$ calculated for a rectangular prism sample~\cite{Aharoni1998_demag}. For measurements with $H\parallel ab$, the sample geometry results in a negligible demagnetizing effect so no correction was applied. 

\section{Pulsed-Field Magnetometry}

Pulsed-field measurements of $M(H)$, are performed in a short-pulse $65$\,T magnet at the National High Magnetic Field Laboratory, Los Alamos. Fields of up to $60$\,T with a  typical rise time of around $10$\,ms were used. Several small single crystals are coaligned and stacked together using GE varnish and placed in PTFE ampoules with an inner diameter of $1.3$\,mm. The demagnetizing effect in the resulting sample geometry is negligible. Measurements are made using a $1.5$\,mm bore, $1.5$\,mm long, 1500-turn compensated coil susceptometer. With the sample in the coil, the signal voltage $V \propto \frac{\mathrm{d}M}{\mathrm{d}t}$, where $t$ is time, is measured. A background measurement with the sample out of the coil and under identical conditions is also performed before integrating $V$ to determine the $M(H)$ curve. The magnetic field is determined via the voltage induced in a 10-turn coil calibrated using the observation of de Haas-van Alphen oscillation of the belly orbits in the copper coils. Finally, the absolute magnetization of the sample is calibrated to the $M(H)$ curves measured using a SQUID magnetometer.

\section{Magnetoresistance}
\begin{figure}[h]
    \centering
    \includegraphics[width= 0.65\linewidth]{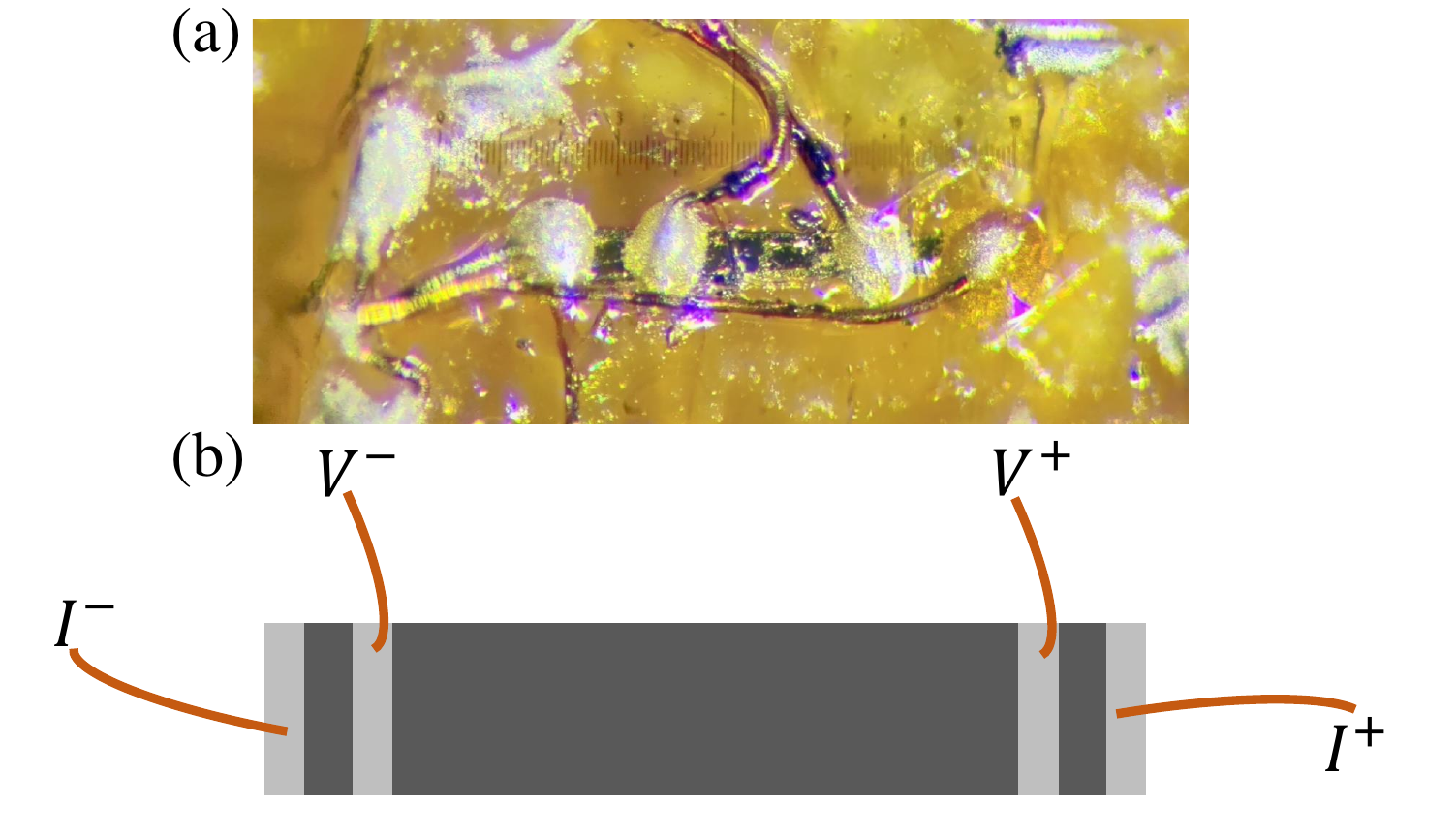}
    \caption{(a) Picture of the sample used for longitudinal magnetoresistance (MR) measurements. The sample was potted in General Electric varnish, to stop movement during pulsed field measurements. (b) A schematic of the four-contact configuration used to measure MR. An alternating current is applied across $I^{-}$ and $I^{+}$ and potential drop across $V^{-}$ and $V^{+}$ is measured.}
    \label{supp_fig:res_setup}
\end{figure}

\begin{figure}[h]
    \centering
    \includegraphics[width= 0.65\linewidth]{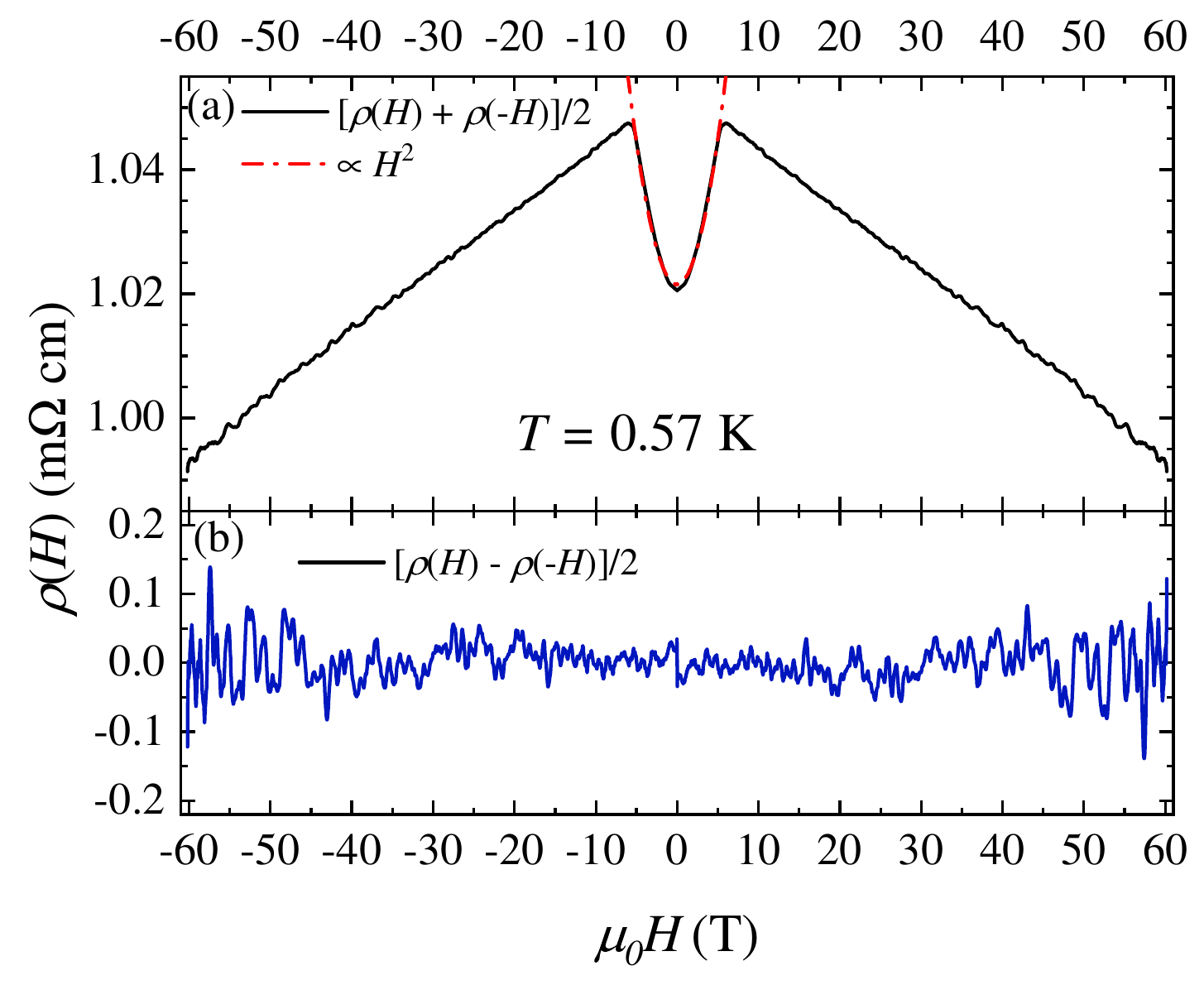}
    \caption{Field dependent magnetoresistance, $\rho(H)$, measured at $0.57$\,K with positive and negative field $H\parallel ab$. (a) The longitudinal resistivity signal (black) that is symmetric with the positive and negative applied fields. The red dashed line serves as a guide to the eye to the symmetric $H^2$ enhancement that is observed below $6$\,T. (b) The negligible anti-symmetric Hall signal which is present in our measurements of $\rho_{xx}$.}
    \label{supp_fig:Sym_MR}
\end{figure} 

\begin{figure} 
    \centering
    \includegraphics[width= 0.65\linewidth]{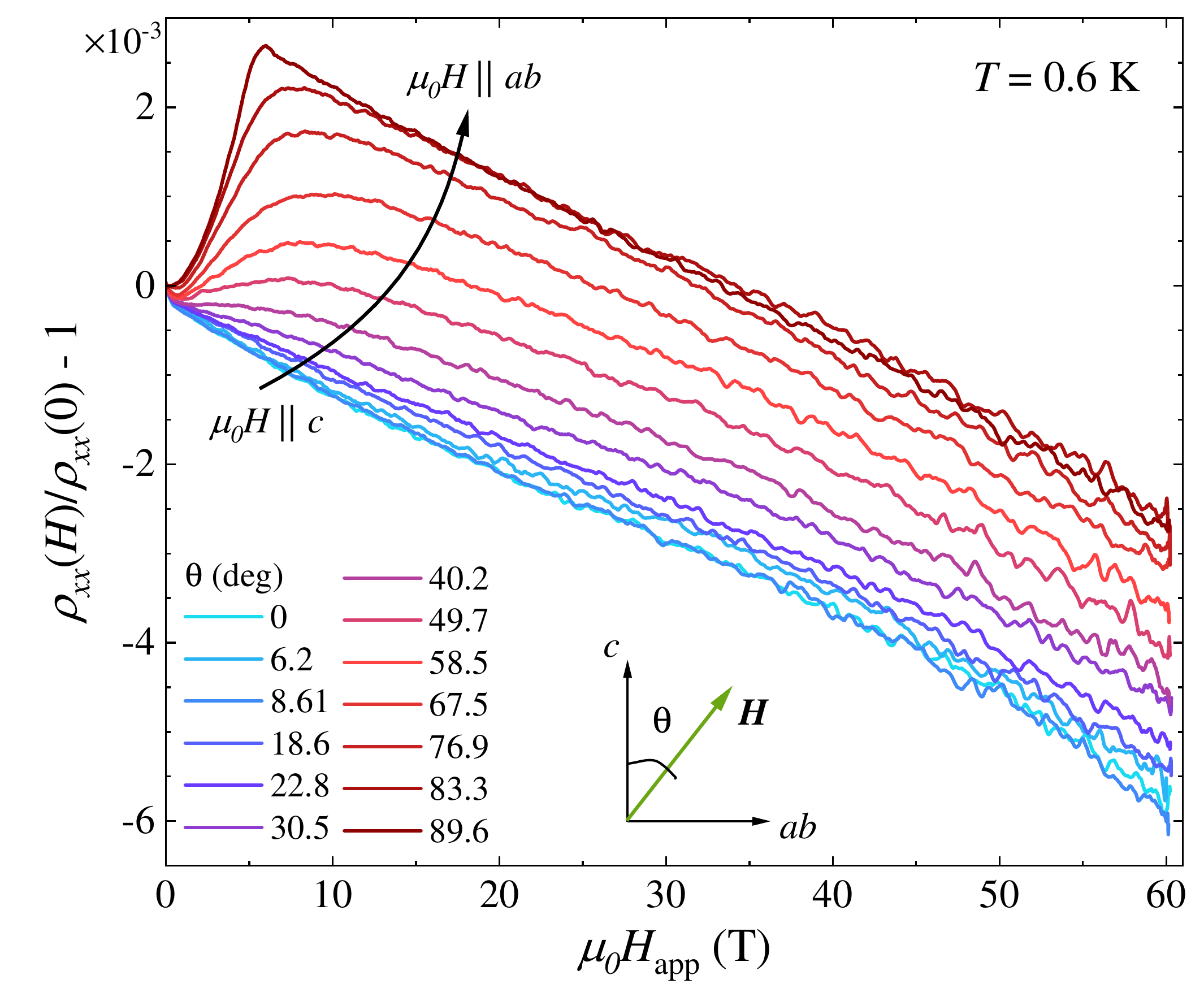}
    \caption{ Magnetoresistance (MR) measurements at $0.6$\,K as the field is rotated from $H\parallel c$ to $H\parallel ab$ with the current $I \perp H$ at all field angles.}
    \label{supp_fig: Angle-dep-MR}
\end{figure}

Pulsed-field measurements, up to $60$\,T, of the longitudinal magnetoresistance (MR) are performed on a rectangular flat-plate sample with dimensions $1.4 \times 0.13$\,mm$^2$ and a height of around $0.04$\,mm along the $c$-axis. A four-contact configuration, as shown in Fig.~\ref{supp_fig:res_setup}, was used, where two contacts supply an alternating current of $I = 1.3(1)$ mA, at a frequency $f= 420$\,kHz, across the length of the sample. Measurements were taken with positive and negative applied field and the longitudinal and Hall components of the resistivity, $\rho(H)$, are then extracted by performing the symmetrizing, $\rho_{xx} = [\rho(H)+\rho(-H)]/2$, and anti-symmetrizing, $\rho_{xy} = [\rho(H)-\rho(-H)]/2$, operations respectively. Fig.~\ref{supp_fig:Sym_MR}(b) shows a negligible Hall component present in measurements of $\rho_{xx}$. The $0.6$\,K MR, presented in Fig.~\ref{supp_fig: Angle-dep-MR}, show that the low-field $H^{2}$ component (discussed in main text) is maximum when $H \parallel ab$.

\begin{figure}[h]
    \centering
    \includegraphics[width= 0.72\linewidth]{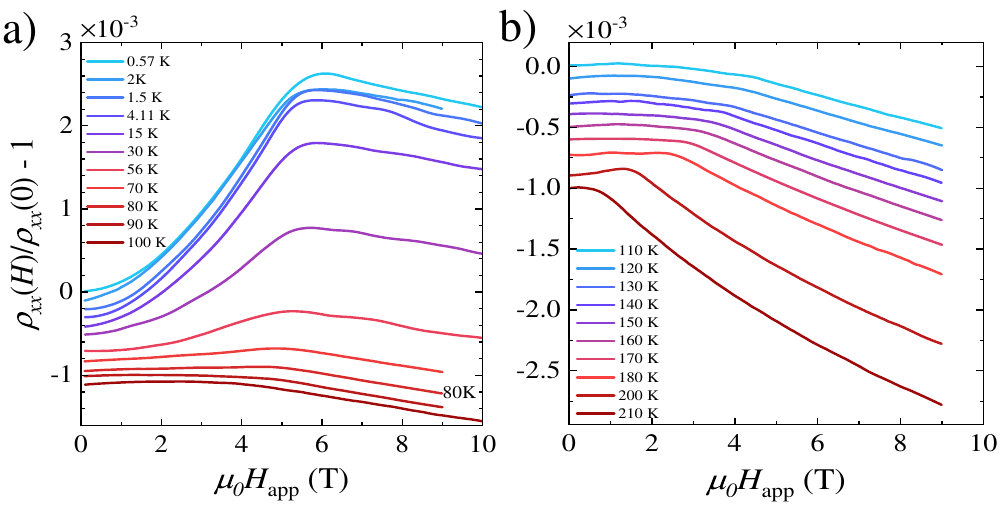}
    \caption{ Magnetoresistance measurements up to $9$\,T with field $H\parallel ab$ at temperatures ranging from (a) $0.57$ to $100$\,K and (b) $110$ to $210$\,K. For clarity, the curves have been offset by $0.1$.}
    \label{supp_fig:MR_ab_Low_Field}
\end{figure}

Additional measurements of the MR with an in-plane field, up to $9$\,T, are performed in a Quantum Design Physical Properties Measurement System (PPMS) with an alternating current of $I = 113$\,mA and $f=113$\,Hz. The low-field $H^{2}$ enhancement is no longer present by $80$\,K, but a shoulder feature persists up to $T_{\text{c}}$, as shown in Fig.~\ref{supp_fig:MR_ab_Low_Field}.

\FloatBarrier

\section{Hall Effect}

\begin{figure}[h]
    \centering
    \includegraphics[width= 0.5\linewidth]{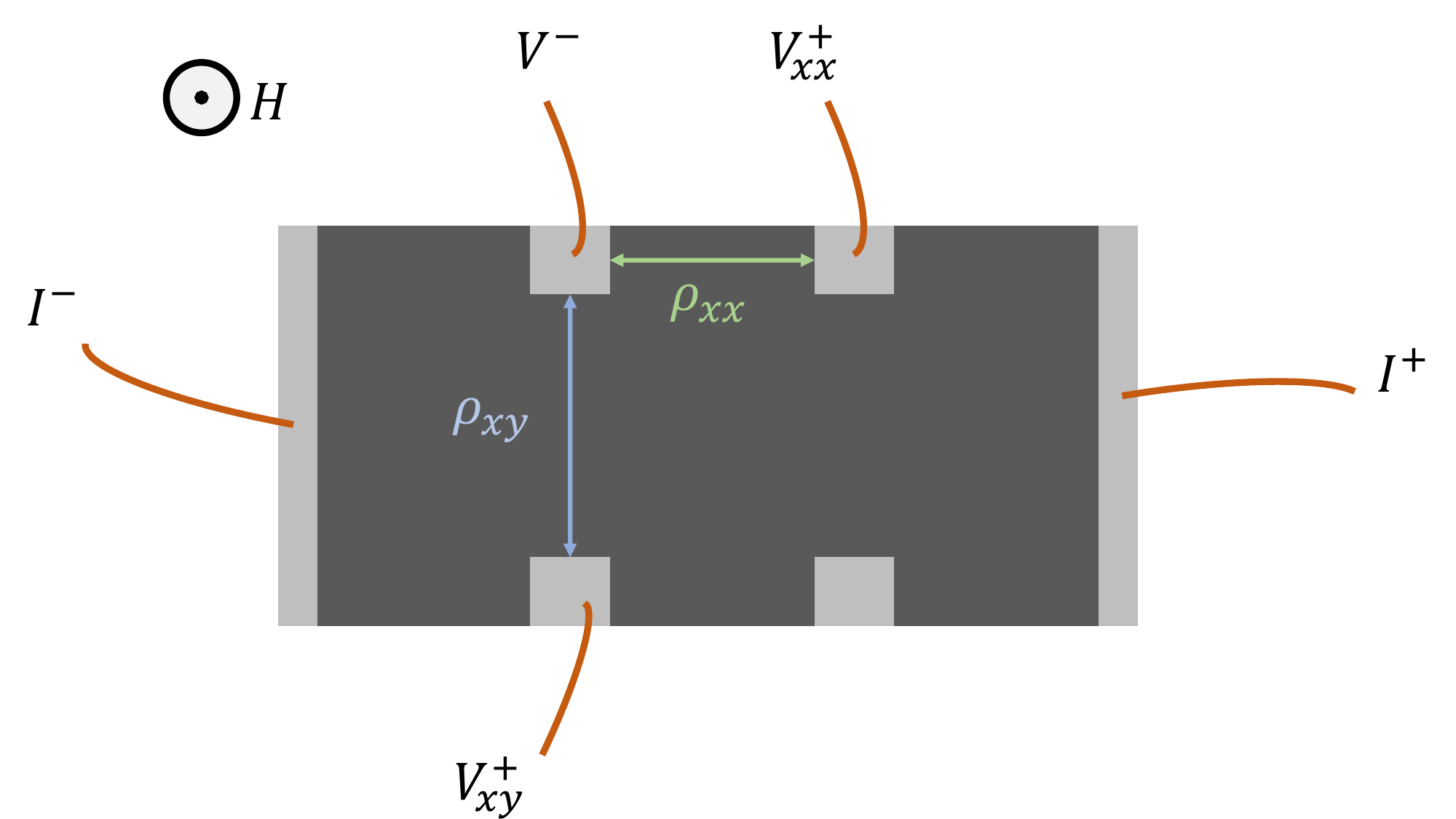}
    \caption{ A schematic of the sample setup used to simultaneously measure $\rho_{xx}$ and $\rho_{xy}$. $\rho_{xy}$ measurements were taken across voltage contacts $V^{+}_{xy}$ and $V^{-}$, while $\rho_{xx}$ measurement were taken across $V^{+}_{xx}$ and $V^{-}$.}
    \label{supp_fig:Hall_setup}
\end{figure}

\begin{figure}[h]
    \centering
    \includegraphics[width= 0.75\linewidth]{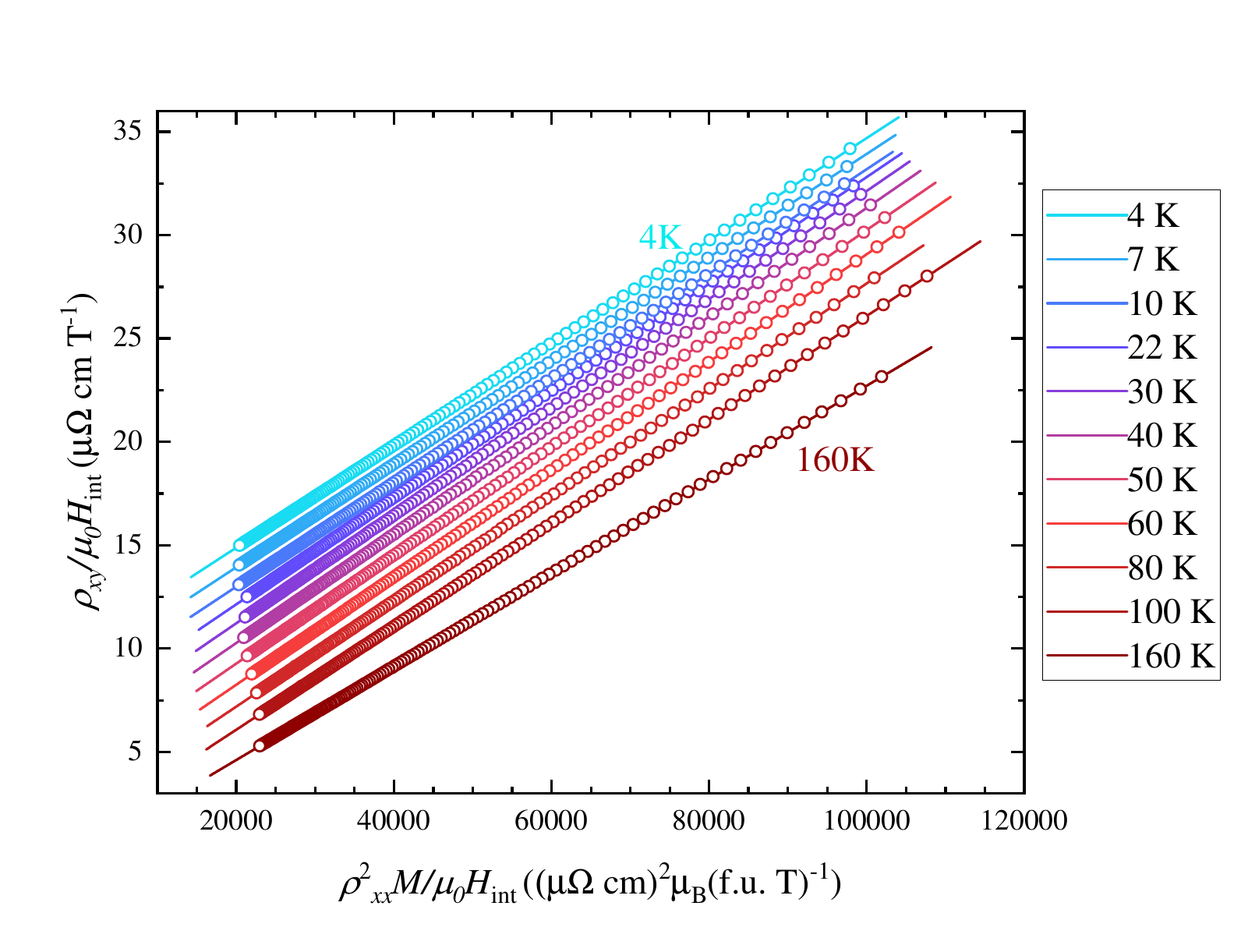}
    \caption{ $\rho_{xy}/\mu_{0}H$ vs $\rho^{2}_{xx}M/\mu_{0}H$ with $\mu_{0}H\parallel c$ and $\geq 1.75$\,T. Data points are displayed as circles and the lines are linear fits to the data at various temperatures. The curves have been offset by $1$\,$\mu\Omega$cmT$^{-1}$ for clarity.}
    \label{supp_fig:KL_line}
\end{figure}

Measurement of $\rho_{xx}$ and $\rho_{xy}$, up to $9$\,T, are performed simultaneously in a PPMS using the configuration shown in Fig.~\ref{supp_fig:Hall_setup}. The $\rho_{xx}$ and $\rho_{xy}$ signals are symmetrized and anti-symmetrized respectively, as described above. The dimensions of the flat-plate-shaped sample used are $2.1 \times 0.8 \times 0.04$\,mm$^3$. As discussed in the main text the Hall resistivity can be modelled as
\begin{equation}
    \label{supp_eq: hall_KL}
    \rho_{xy} = R_{0}\mu_{0}H + S_{\text{H}}\rho^{2}_{xx}M,
\end{equation}
where $S_{\text{H}}$ is a scaling factor for the anomalous Hall effect (AHE) and $R_{0}$ is the ordinary Hall coefficient~\cite{Wang2017, kim2018, Saha2023}. Both of these parameters can be extracted from linear fits to $\rho_{xy}/\mu_{0}H$ vs $\rho^{2}_{xx}M/\mu_{0}H$, for $\mu_{0}H \geq 1.75$\,T, as shown in Fig.~\ref{supp_fig:KL_line}. 

\begin{figure}
    \centering
    \includegraphics[width= \linewidth]{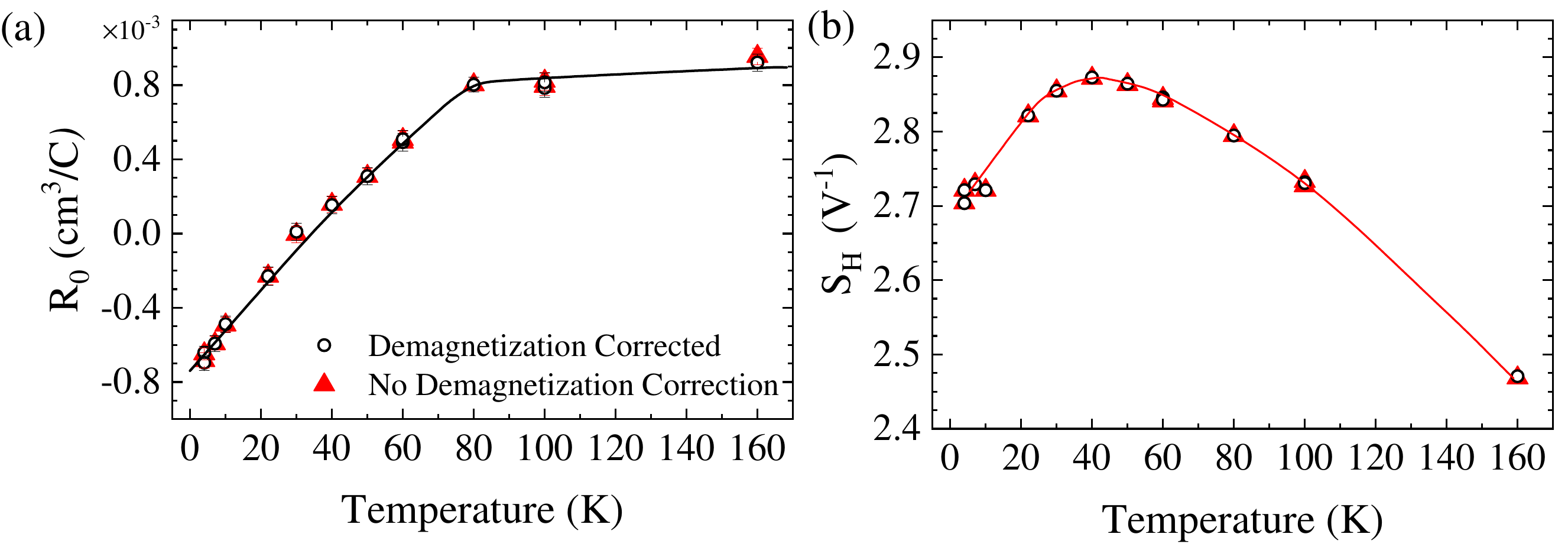}
    \caption{Temperature dependence of (a) ordinary Hall coefficient, $R_{0}$, and (b) the anomalous Hall effect scale factor $S_{\text{H}}$. In both, values depicted with black circles are corrected for demagnetization effects, while the values indicated by the red triangles are not.}
    \label{supp_fig:R0_demag_v_no-demag}
\end{figure}

Due to the flat-plate geometry of the sample and the differences in the dimensions of the magnetometry and transport samples, demagnetization corrections had to be applied for the $H\parallel c$ measurements. The resistivity as a function of the internal field $\rho_{xy}(H)$ was calculated using
\begin{equation}
    \label{supp_eq: rho_H_int}
    \rho_{xy}(H) = \rho_{xy}[H_{\text{app}} - D_{\rho}M_{\rho}(H_{\text{app}})],
\end{equation}
where the demagnetizing factor $D_{\rho} = 0.909$. $M_{\rho}(H_{\text{app}})$ is the expected magnetization as a function  of $H_{\text{app}}$ for the transport sample and is calculated using,
\begin{equation}
    \label{eq/mag_rho_int}
    M_{\rho}(H_{\text{app}}) = M[H + D_{\rho}M(H)].
\end{equation}
This assumes that there is negligible sample dependence of the $M(H)$ curves, which is expected as the samples are from the same batch. We note that applying demagnetization correction in fact has a near negligible effect on the extracted $R_{0}$ and $S_{\text{H}}$ values (see Fig.~\ref{supp_fig:R0_demag_v_no-demag}).

\begin{figure}[h]
    \centering
    \includegraphics[width= 0.75\linewidth]{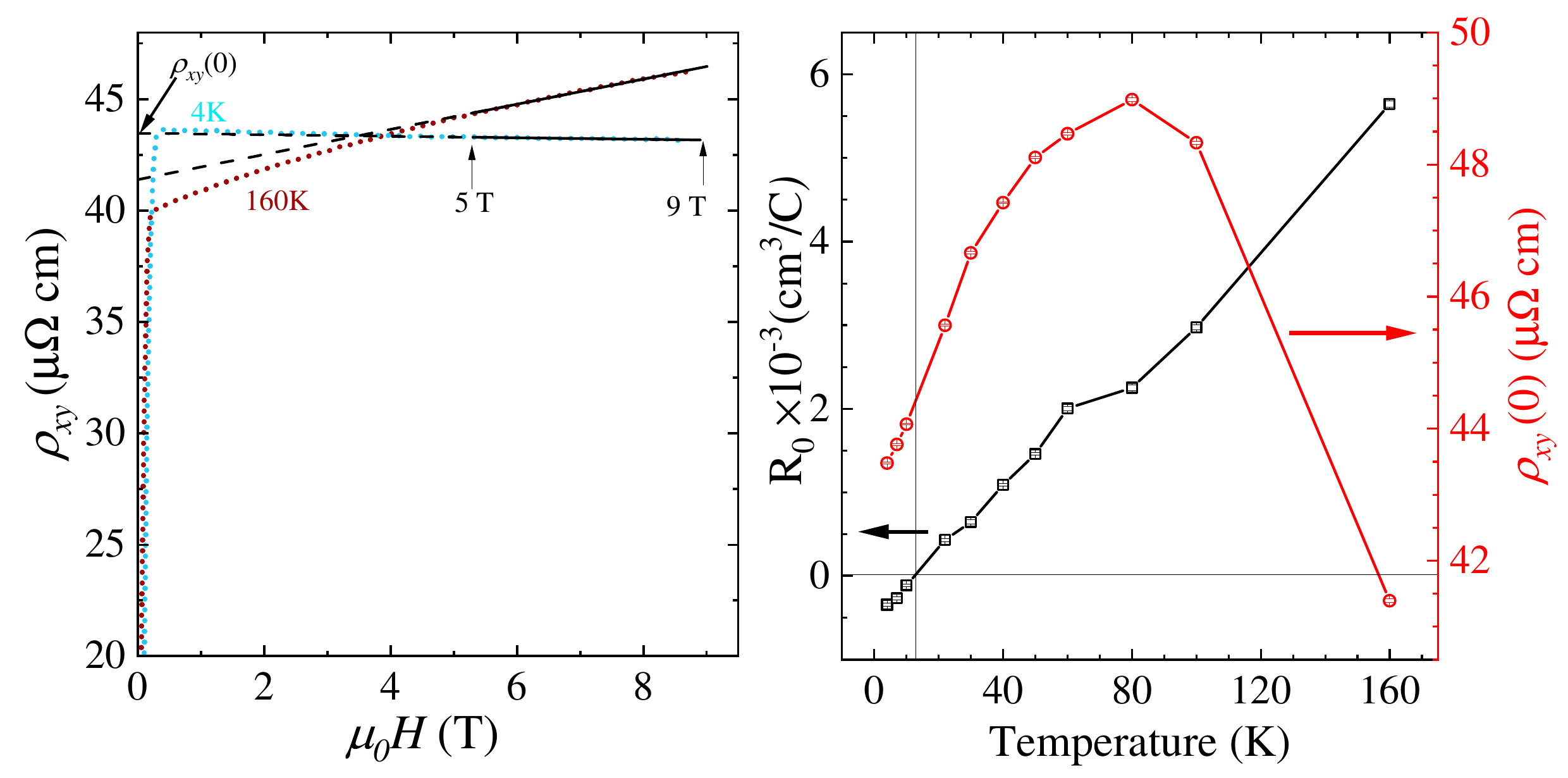}
    \caption{Analysis of Hall effect assuming zero field dependence of the AHE. (a) Hall resistivity, $\rho_{xy}$, at $4$\,K (blue dashed line) and $160$\,K (red dashed line). The solid black lines show the linear fit to $\rho_{xy}$, for $\mu_{0}H\geq5$, and the dashed black lines depict an extrapolation to $0$\,T. (b) Temperature dependence of the ordinary Hall coefficient $R_{0}$ (black squares) and $\rho_{xy}(0)$ (red squares). This analysis produces erroneous results, as described in the text.} 
    \label{supp_fig:R0_wang}
\end{figure}

Previous analyses of the Hall effect in FGT, have neglected the significant field dependence of the AHE arising from $M(H)$ and $\rho_{xx}(H)$~\cite{Wang2017, Saha2023}. In this assumption, the second term in Eq.~\ref{supp_eq: hall_KL} is a constant and $R_{0}$ is obtained through linear fits to $\rho_{xx}$ for $\mu_{0}H\geq5$\,T, as shown in Fig.~\ref{supp_fig:R0_wang}(a). If we analyze our data in this way, the resulting $R_{0}$, shown in Fig.~\ref{supp_fig:R0_wang}(b), depicts a decreasing trend on cooling similar to Ref.~\cite{Wang2017} but with a change in sign for $T<15$\,K. This analysis--which assumes that the AHE has no field dependence--misses out on the electronic transition occurring at $80$\,K and crossover of majority carrier at around $35$\,K. As we have discussed in the main text, there is a significant field-dependent contribution  to the AHE and this must be considered in order to reliably extract $R_{0}$.

\FloatBarrier
\bibliography{supp.bib}